
\input harvmac
\def\st{space-time}
\def\Pmm{\Pi^\mu_{\mu'}}\def\Pnn{\Pi^\nu_{\nu'}}
\def\gt{\widetilde\nabla}
\def\Fumn{F^{\mu\nu}}\def\Flmn{F_{\mu\nu}}
\def\flmn{f_{\mu\nu}}
\def\qv{\textstyle{1\over qv}}
\def\IR{{\rm I}\kern-.20em{\rm R}}
\lref\no{H. B. Nielsen and P. Olesen,  Nucl. Phys. {\bf 61B}, 45 (1973).}
\lref\wald{ R. M. Wald, {\it General Relativity}, Univ. Chicago Press, Chicago,
1984, and references therein.}
\lref\cpwa{S. Coleman, J. Preskill and F. Wilczek,  Phys. Rev. Lett. {\bf
67} 1975 (1991).}
\lref\cpwb{S. Coleman, J. Preskill and F. Wilczek,  Mod. Phys. Lett. A
{\bf 6}, 1631 (1991).}
\lref\gib{G. W. Gibbons, {\it Self gravitating Magnetic Monopoles, Global
Monopoles and Black Holes}, 12th Lisbon Autumn School on Physics,
Springer-Verlag, 1991.}
\lref\gihaw{G. W. Gibbons and S. W. Hawking,
Phys. Rev. D {\bf 15}, 2752 (1977).}
\lref\adP{S. Adler and R. Pearson, Phys. Rev. D {\bf 18}, 2798 (1978).}
\lref\bghhs{M. J. Bowick, S. B. Giddings, J. A. Harvey, G. T. Horowitz and
A. Strominger,  Phys. Rev. Lett. {\bf 61} 2823 (1988).}
\lref\abl{T. J. Allen, M. J. Bowick and A. Lahiri, Phys. Lett. {\bf
237B} 47 (1989).}
\lref\cha{J. E. Chase, Comm. Math. Phys. {\bf 19}, 276 (1970).}
\lref\bek{J. D. Bekenstein, Phys. Rev. D{\bf 5}  1239 (1972); {\bf 5},
2403 (1972).}
\lref\beki{J. D. Bekenstein, Ann. Phys. (N. Y.), {\bf 91}, 75 (1975);
N. Bocharova, K. Bronnikov and V. Melnikov, Vestn. Mosk. Univ. Fiz. Ast.
{\bf 6}, 706 (1970).}
\lref\kw{L. M. Krauss and F. Wilczek, Phys. Rev. Lett. {\bf 62}, 1221 (1989).}
\lref\jmrw{J. Minahan and R. Warner, {\sl Stuckelberg Revisited},
University of Florida preprint UFIFT-HEP-89-15.}
\lref\abla{T. J. Allen, M. J. Bowick and A. Lahiri, Mod. Phys. Lett. A {\bf 6}
(1991) 559.}
\noblackbox
\Title{\vbox{\baselineskip12pt\hbox{LA-UR-92-1861}\hbox{gr-qc/9207008}}}
{The No-Hair Theorem for the Abelian Higgs Model}
\centerline{Amitabha Lahiri\footnote{$^{\dag}$}{(lahiri@pion.lanl.gov)}}
\bigskip\centerline{Theoretical  Division  T-8}
\centerline{Los Alamos National Laboratory}
\centerline{Los Alamos, NM 87545, USA}
\vskip0.3in
\centerline{\bf Abstract}
\noindent We consider the general procedure for proving no-hair theorems
 for static, spherically symmetric
black holes. We apply this method to the abelian Higgs model and find a
proof of the no-hair conjecture that circumvents the objections raised against
the original proof due to Adler and Pearson.
\Date{06/92}

\newsec{Introduction}

Classical ``no-hair'' theorems \wald\ state that a stationary black hole is
characterized by a small number of parameters -- its mass, angular momentum,
and charges corresponding to long-range gauge fields. It has long been known
 \refs{\cha, \bek} that static
black holes carry no hair, or external field, corresponding to massless or
massive scalar fields or Proca-massive gauge fields.  It was
also shown
sometime later \adP\ that if an abelian gauge field acquires a mass via the
Higgs mechanism, the corresponding gauge field must vanish outside the horizon.
However, the arguments used there have recently been criticized \gib\ as being
too restrictive. Therefore it is necessary to find a more rigorous proof of the
no-hair conjecture before one can believe in it.

On the other hand, a failure to find such a proof, or more precisely,
evidence that a rigorous proof cannot exist, will have very interesting
consequences. It has been recognized \refs{\bghhs,\kw} that the classical
no-hair theorems do not rule out the possibility that black holes may carry
charges that are detectable only through experiments of a quantum nature.
The $Z_N$ quantum hair as it is called \kw\ arises from spontaneous
symmetry breaking in an abelian Higgs model, where the Higgs condensate has
charge $N\hbar e$, $\hbar e$ being the charge quantum of the theory. The
dynamical effect of this quantum hair is expected to be non-perturbative in
$\hbar$, with a serious effect on the thermodynamics of a static black hole
\refs{\cpwb, \cpwa}.
\nref\cpw{S. Coleman, J. Preskill and F. Wilczek, {\it Quantum Hair on
Black Holes}, CalTech preprint IASSNS-HEP-91-64, 1991.}\nref\dgt{F. Dowker, R.
Gregory and J. Traschen, Phys. Rev. D {\bf 45} 2762 (1992).}
The partition function of a black hole at temperature
$\beta^{-1}$ is given by \gihaw\
 the path-integral of the Euclidean action
over asymptotically flat, topologically $R^2\times S^2$ configurations that
are periodic in the imaginary time $\tau$ with period $\beta\hbar$. The
saddle points of this path-integral include classical Euclidean black hole
solutions coupled to non-trivial gauge and Higgs field configurations.
It was conjectured \refs{\cpwb, \cpwa} that such solutions
indeed exist, and they behave like vortices on the $\tau-r$ plane on an
`almost Schwarzschild' Euclidean background. More precise calculations
followed \refs{\cpw, \dgt}, and even though no exact solution was found,
stronger arguments were given for their existence. If the
no-hair conjecture for the abelian Higgs model is found to be incorrect,
one will have
found more evidence for the existence of these solutions. On the other hand,
if it is correct, one has to demonstrate the failure of the proof for
Euclidean backgrounds in order for these `Euclidean' vortices to exist.

Another reason for looking for a rigorous proof of the no-hair conjecture
is the following. It was found in \bghhs\ that a static black hole can
carry a topological charge corresponding to the surface integral of an
antisymmetric tensor potential, and later it was shown \abl\ that this
special `hair' persists even when this potential becomes massive via a
coupling to a massless abelian gauge field. The action used for this
purpose can also have an interpretation by which the gauge field becomes
massive after absorbing the degree of freedom in the tensor potential
\refs{\jmrw, \abla}. Thus, contrary to prevalent belief, a black hole $can$
carry some information apart
from its mass and angular momentum in the presence of a gauge field that
acquires a gauge-invariant mass. Even though this does not imply the
failure of the no-hair conjecture for the abelian
Higgs model,
it does raise some skepticism.

In light of the various results mentioned above, we propose to take a
renewed look at the classical no-hair theorems. We consider the general
procedure used to prove such theorems, specifying all the assumptions that
are used. We apply this procedure to the abelian Higgs model coupled to
gravity and find a rigorous proof for the no-hair conjecture. This proof
fails when the space-time metric has a Euclidean signature, corroborating
the arguments of \refs{\cpwb-\dgt}.

\newsec{General Setup}

We restrict ourselves to a $3 + 1$-dimensional static, spherically symmetric,
asymptotically flat \st\ with a horizon. This implies making the following
assumptions: \hfill\break
($i$) The \st\ is endowed with a timelike Killing vector $\xi^\mu$ with
$\xi^\mu\xi_\mu = - \lambda^2(r)$ which obeys $\xi_{[\mu}\grad\nu\xi_{\lambda]}
= 0$; (it follows \wald\ that there is a spacelike hypersurface $\Sigma$ ---
the space `outside' the horizon --- which is everywhere orthogonal to
$\xi^\mu$.) \hfill\break
($ii$) The hypersurface $\Sigma$ allows a coordinatization isomorphic to the
flat-space spherical coordinates (the space-time metric may be written as
$ds^2 = -\lambda^2(r)dt^2 + h^2(r)dr^2 + r^2d\Omega$); \hfill\break
($iii$) $\lambda$ vanishes at a finite value $r_H$ of the radial coordinate
$r$, thus defining the horizon;\hfill\break
($iv$) $\lambda \to 1 + O(1/r)$ as $r \to \infty$ (asymptotic flatness).
\hfill\break
These are all the assumptions we will need to make about the \st, now we turn
our attention to the fields that live on this \st. The crucial assumption that
goes into proving the standard no-hair theorems is that the squared norm of the
stress-energy tensor is
bounded at the horizon and vanishes suitably rapidly at infinity
\refs{\cha - \adP}. This may be seen as being dictated by Einstein's equations.
If the stress-energy tensor $T_{\mu\nu}$ has unbounded norm at any point, the
Einstein tensor and therefore the curvature must also become unbounded there,
giving rise to a singularity. The horizon is not, however, a a curvature
singularity, but only a coordinate singularity. Therefore the stress-energy
tensor must remain bounded at the horizon. Similarly, asymptotic flatness
dictates that the metric approaches the Schwarzschild metric as $r\to \infty$.
It follows that $T_{\mu\nu}$ must vanish in this region. Similar arguments show
that $T_{\mu\nu}$ must also be static, {\it i.e.}, have vanishing Lie
derivative with respect to $\xi^\mu$.

We will need one more result for proving no-hair theorems, which we
write down here.
Let us denote the projection operator that projects down to $\Sigma$ by $\Pmm
:= \delta^\mu_{\mu'} + \lambda^{-2}\xi^\mu\xi_{\mu'}$. Let us also denote the
\st\ connection by $\grad\mu$ and the induced connection on $\Sigma$ by
$\gt_\mu$. Then for a rank $p$ antisymmetric tensor $\Omega$ whose
$\Sigma$-projection is $\omega$ and $\biglie_\xi\Omega = 0$, it can be shown
that\foot{See Appendix A.}
\eqn\divtheo{\gt_\alpha(\lambda\omega^{\alpha\mu\cdots\nu}) =
\lambda\grad\alpha\Omega^{\alpha\mu'\cdots\nu'}\Pmm\cdots\Pi^\nu_{\nu'}.}
Physically this may be understood as the statement that the 4-divergence of
$\Omega$ is equal to its 3-divergence when the metric and $\omega$ are
time-independent.

The algorithm for proving no-hair theorems may be seen in our first example of
a real scalar field $\rho$ moving in a potential $U(\rho)$. The Lagrangian is
\eqn\rholag{\CL = -(\half\grad\mu\rho\nabla^\mu\rho + U(\rho)),}
and the equations of motion are
\eqn\rhoeom{\grad\mu\nabla^\mu\rho = {\del U\over\del\rho} \equiv\ U'(\rho).}
Using the divergence relation \divtheo\ we can write
\eqn\rhoeoma{\gt_\mu(\lambda\gt^\mu\rho) = \lambda U'(\rho).}
Multiplying both sides by $\rho$ and integrating over the space-like region
$\Sigma$
between the horizon and infinity, we get
\eqn\baldrho{\int_{\del \Sigma}\lambda\rho\gt_\mu\rho n^\mu -
\int_\Sigma\lambda(\gt_\mu\rho\gt^\mu\rho + \rho U'(\rho)) = 0,}
where $\del\Sigma$ is composed of the spheres at the horizon and at infinity,
and
$n^\mu$ is the outward pointing space-like unit normal on these two spheres.
$\gt_\mu\rho\gt^\mu\rho$ appears in $T_{\mu\nu}$, so must be bounded at the
horizon and vanish at infinity. Since
the metric on $\Sigma$ is positive definite, we may apply Schwarz inequality,
which says $|\gt_\mu\rho n^\mu|^2 \leq (\gt_\mu\rho\gt^\mu\rho)(n_\mu n^\mu)
= (\gt_\mu\rho\gt^\mu\rho)$, since $n^\mu$ is a unit vector. It follows that
$\gt_\mu\rho n^\mu$ has to obey the said boundedness conditions. If
$\rho$ is massive ($U(\rho) = \half m^2\rho^2$), the behavior of $T_{\mu\nu}$
also dictates
that $\rho$ has to be bounded everywhere on or outside the horizon. It follows
that the integral over the boundary
$\del \Sigma$
vanishes\foot{For the massless scalar, this is enforced by demanding that
$\rho$ remains measurable and thus bounded at the horizon \refs{\cha, \bek}.},
the
volume integral is an integral of a sum of squares (the metric on
$\Sigma$ is positive definite) and therefore $\rho$ must be trivial outside the
horizon. The same result also holds when $U
= \alpha\rho^4$ with $\alpha > 0$ (or any other convex potential). The
situation is different when $U$
is a double-well potential
$U = {\alpha\over 4}(\rho^2 - v^2)^2$. There is no known rigorous proof of
the no-hair conjecture for the scalar field in such a potential \gib.

({\sl Note:} If $\rho$ is a conformal scalar field, $U = {1\over 12}R\rho^2$,
one cannot impose boundedness conditions on $\rho$ because of the local
conformal
symmetry of the theory. It follows that the boundary integrals need not
vanish. Thus even
if we are looking for solutions with $R = 0$, we can find a non-trivial
conformal scalar field outside the horizon \beki.)

\newsec{The Abelian Higgs Model}

Now we are ready to look for a proof of the no-hair conjecture in the case of
the Abelian Higgs model.
We will work with the Lagrangian
\eqn\HiggsL{\CL = - ({\textstyle{1\over 4}}\Flmn\Fumn +
\half(D_\mu\Phi)^*D^\mu\Phi + U(\Phi)).}
Here $\Phi$ is a complex scalar field, $\Flmn = \grad{[\mu}A_{\nu]}$ is the
field strength of the Abelian gauge
field $A_\mu$, $D_\mu\Phi = (\grad\mu + iqA_\mu)\Phi$ is the gauge covariant
derivative, and $U(\Phi) = {\alpha\over 4}(|\Phi|^2 - v^2)^2, \alpha > 0,$ is
the Higgs potential.
If we parametrize $\Phi$ as $\Phi = \rho e^{i\eta/v}$, we can see that the
Lagrangian is left invariant by gauge transformations $A_\mu \to A_\mu
+ \grad\mu\chi, \eta \to \eta - ivq\chi$.
In terms of $\rho$ and $\eta$, the
Lagrangian reads
\eqn\HiggsLb{\CL = -({\textstyle{1\over 4}}\Flmn\Fumn + \half\grad\mu\rho
\nabla^\mu\rho + \half\rho^2q^2(A_\mu + \qv\grad\mu\eta)(A^\mu +
\qv\nabla^\mu\eta) + {\textstyle{\alpha\over 4}}(\rho^2 - v^2)^2).}
 One of the objections raised in \gib\ about the proof
given in \adP\ was that there may not be any non-singular gauge choice in which
both $A_\mu$ and $\Phi$ are static and/or have bounded norm at the horizon.
However, the squared norms of the temporal and spatial components of the
combination $(A_\mu + \qv\grad\mu\eta)$ appear in $T_{\mu\nu}$. Therefore,  we
can take $(A_\mu + \qv\grad\mu\eta)$ to be static as well as of bounded norm at
the horizon and vanishing norm
at infinity.

Let us denote the $\Sigma$-projections of $A_\mu$ and $\Flmn$ by $a_\mu$
and $\flmn$, respectively. The electric potential $\phi :=
\lambda\inv\xi^\alpha A_\alpha$
and the electric field $e^\mu :=
\lambda\inv\xi_\alpha F^{\mu\alpha}$ satisfy the relations
\eqn\eleceom{\gt_\mu e^\mu = \lambda\inv\xi_\alpha\grad\mu F^{\mu\alpha},\
\ \ \
\gt_\mu{(\lambda\phi)} = \lambda e_\mu +
\biglie_\xi A_\mu.}
 We
follow the procedure used for
a real scalar field, first writing down the
equations of motion for the $\Sigma$-projections of the fields plus the one
coming
from \eleceom, multiply by the appropriate fields and integrate over the region
between the horizon and infinity.
We concentrate on one of the field equations,
\eqn\divelec{\gt_\mu e^\mu - q^2\rho^2(\phi +
\qv\lambda\inv\dot\eta) = 0,}
where we have written $\xi^\mu\grad\mu\eta$ as $\dot\eta$.
We multiply this equation by $\lambda(\phi +
\qv\lambda\inv\dot\eta)$ and integrate over the region between the
horizon and infinity. Including all terms, the resulting equation is
\eqn\intelec{\eqalign{\int_{\del \Sigma}\lambda(\phi +&
\qv\lambda\inv\dot\eta)e^\mu n_\mu \hfill\cr &-
\int_\Sigma(\lambda e^\mu e_\mu + e^\mu\biglie_\xi(A_\mu + \qv\grad\mu\eta) +
\lambda q^2\rho^2(\phi + \qv\lambda\inv\dot\eta)^2) = 0.\hfill\cr}}
Since the squared norms of the spatial and temporal components of  $(A_\mu
+ \qv\grad\mu\eta)$ appear in $T_{\mu\nu}$, this quantity must be static.
Therefore the second term in the volume integral vanishes. The remaining
terms of the integrand are positive indefinite, and must vanish if the
surface integral is zero. The stress tensor $T_{\mu\nu}$ contains the product
$\rho^2(\phi + \qv\lambda\inv\dot\eta)^2$, and the boundary conditions on
$T_{\mu\nu}$ imply
the following. Since $\rho\to v$ as $r\to\infty$, we must have $(\phi +
\qv\lambda\inv\dot\eta) \to 0$ (as well as $e^\mu n_\mu \to 0$) as $r
\to\infty$, and therefore the contribution to the surface integral
from the sphere at infinity must vanish. The contribution from the horizon
may be non-zero, but then it follows from the boundedness of $T_{\mu\nu}$
that $\rho\to 0$ as $r\to r_H$. The remainder of the proof consists of
showing that $\rho \neq 0$ at $r = r_H$.

The $\Sigma$-projection of the equation of motion for $\rho$ is
\eqn\rhoeomb{\gt_\mu\lambda\gt^\mu\rho = \lambda\alpha\rho(\rho^2 - v^2)
- \lambda q^2\rho(\phi + \qv\lambda\inv\dot\eta)^2,}
assuming for the moment that $(a_\mu + \qv\gt_\mu\eta) = 0$. If $\rho = 0$
 at $r = r_H$ and $\rho\to v$ as $r\to \infty$,
there are only the following possibilities:
\item{$(i)$} $v\geq\rho\geq 0$ for all $r\geq r_H$, $\rho\to v$ monotonically
as $r\to\infty$;
\item{$(ii)$} $v\geq\rho\geq 0$ for $r_v\geq r\geq r_H$, $\rho = v$ at $r =
r_v$;
\item{$(iii)$} $v\geq\rho\geq 0$ for $r_{max}\geq r\geq r_H$, $\rho$ has
a local maximum at $r = r_{max}$ ($n^\mu \gt_\mu\rho|_{r_{max}} = 0$);
\item{$(iv)$} $-v\leq\rho\leq 0$ for $r_{-v}\geq r\geq r_H$, $\rho = -v$
at $r = r_{-v}$;
\item{$(v)$} $-v\leq\rho\leq 0$ for $r_{min}\geq r\geq r_H$, $\rho$ has a
local minimum at $r = r_{min}$ ($n^\mu\gt_\mu\rho|_{r_{min}} = 0$).

For cases $(i) - (iii)$, we multiply \rhoeomb\ by $(\rho - v)$ and integrate
over a region $V$ between the horizon and the spheres respectively at
$(i)$ infinity, $(ii)\ r_v$, and $(iii)\ r_{max}$. The resulting equation
is
\eqn\nohair{\int_{\del V}\lambda(\rho - v)n^\mu\gt_\mu\rho
- \int_V\lambda(\gt_\mu\rho\gt^\mu\rho + (\rho - v)^2\rho(\rho + v)
- \rho(\rho - v)(\phi + \qv\lambda\inv\dot\eta)^2) = 0.}
By our choice of the region $V$, the surface integral over $\del V$ vanishes
in each case ($\lambda = 0$ and $\rho, n^\mu\gt_\mu\rho < \infty$
at the horizon), and since $v\geq\rho\geq0$ everywhere on $V$, the integrand
is necessarily
positive definite, {\it i.e.}, we have a contradiction. For the cases $(iv)$
and $(v)$,
we multiply \rhoeomb\ by $(\rho + v)$ and integrate over the region between
the horizon and $(iv)\ r_{-v}$, $(v)\ r_{min}$, respectively. Again we find
that the integral of a positive definite quantity must be zero. Since the
cases $(i) - (v)$ exhaust the possible behaviors of $\rho$ if it has to vanish
at the horizon and reach $v$ at infinity, we conclude that there is no
such solution. It follows that the surface integral in \intelec\ must vanish
as well (finiteness of $T_{\mu\nu}$ and the non-vanishing of $\rho$ demands
the finiteness of $(\phi + \qv\lambda\inv\dot\eta)$ at the horizon), and
the black hole carries no gauge hair.

We also need to justify the assumption that $(a_\mu + \qv\gt_\mu\eta)
= 0$, and
the justification is the following. The equation of motion for $a_\mu$
leads to the integral
\eqn\nohaira{\int_{\del\Sigma}\lambda(a_\mu + \qv\gt_\mu\eta)f^{\mu\nu}n_\nu
- \int_\Sigma\lambda(\half f^{\mu\nu}f_{\mu\nu} + q^2\rho^2(a_\mu +
\qv\gt_\mu\eta)^2) = 0.}
Obviously, the only way to have $(a_\mu + \qv\gt_\mu\eta)$ non-vanishing
outside the  horizon is to allow it to diverge at least as fast as
$\lambda\inv$ at the horizon. But the $f_{\mu\nu} = \gt_{[\mu}a_{\nu]}$ will
diverge as fast as $\lambda^{-2}$ and $f^{\mu\nu}f_{\mu\nu}$, a quantity
appearing in $T_{\mu\nu}$, will also diverge. (The appearance, in a
coordinate basis, of $g^{rr}$ in
$f^{\mu\nu}f_{\mu\nu}$ cannot nullify this divergence because $\sqrt{g^{rr}}$
appears in the surface integral as well. Then, assuming $g^{rr}$ vanishes
faster
than $\lambda^2$, we find that $(a_\mu + \qv\gt_\mu\eta)$ has to diverge as
$1/(\lambda\sqrt{g^{rr}})$, {\it i.e.}, $f^{\mu\nu}f_{\mu\nu}$ diverges as
$1/(\lambda^2g^{rr})$.)
Since this is contradictory to our assumption that $T_{\mu\nu}$ is bounded at
the horizon, we conclude that  $(a_\mu + \qv\gt_\mu\eta)$ cannot diverge
at the horizon and therefore must vanish everywhere on $\Sigma$ according to
\nohaira.

Thus we have proven the no-hair conjecture for the abelian Higgs model
coupled to a static, spherically symmetric black hole without making the
restrictive assumptions made in the original proof. This proof is gauge-
invariant, so the objections raised in \gib\ have been taken care of.

Finally, the above proof of the no-hair conjecture for the abelian
Higgs model
does not apply to a metric with Euclidean signature. The equations
\intelec\ and \nohaira\ hold, as well as the arguments following them,
but the equation \rhoeomb\ is now different (the last term changes sign),
and the arguments following it do not hold any more.

\centerline{\bf Acknowledgements}

It is pleasure to thank T. Allen, F. Dowker, B. Greene and E. Mottola for
stimulating discussions.

\appendix{A}{Derivation of \divtheo}

Frobenius' condition for hypersurface-orthogonality for a Killing vector
$\xi^\mu$ states that \wald\ $\xi_{[\mu}\grad\nu\xi_{\lambda]} = 0$.
Contracting with $\xi^\lambda$ it follows that
\eqn\xider{\Pmm\Pnn\grad\mu\xi_\nu = 0.}
Also, from the definition of $\lambda$,
\eqn\lamder{\gt_\mu\lambda = \Pi^{\mu'}_\mu\grad{\mu'}\lambda \equiv
\grad\mu\lambda = -\lambda^{-1}\xi_\alpha\grad\mu\xi^\alpha.}
Now we are ready to look at the left hand side of \divtheo,
\eqn\lhslem{\eqalign {\gt_\alpha(\lambda\omega^{\alpha\mu\cdots\nu}) &= \lambda
\gt_\alpha\omega^{\alpha\mu\cdots\nu} + \omega^{\alpha\mu\cdots\nu}
\gt_\alpha\lambda\cr
&=
\lambda\Pi^\alpha_{\alpha'}\Pmm\cdots\Pnn\grad\alpha
\Omega^{\alpha'\mu'\cdots\nu'} +
  \omega^{\alpha\mu\cdots\nu}\gt_\alpha\lambda\cr
&= \lambda\grad\alpha\Omega^{\alpha\mu'\cdots\nu'}\Pmm\cdots\Pnn +
\lambda\inv\xi^\alpha\xi_{\alpha'}\grad\alpha
\Omega^{\alpha'\mu'\cdots\nu'}\Pmm\cdots\Pnn \cr
&\qquad\qquad +
\Pi^\alpha_{\alpha'}\Pmm\cdots\Pnn(-\lambda\inv\xi_\rho\grad\alpha\xi^\rho)
\Omega^{\alpha'\mu'\cdots\nu'}.\cr}}

By our assumptions, $\biglie_\xi\Omega = 0$ (this may be thought of as the
statement that $\Omega$ is time-independent if one sets up a system of
coordinates where time is parametrized along $\xi$). Thus we can make a
substitution for the second term of the last line of the above equation,
\eqn\lhssub{\eqalign{\gt_\alpha(\lambda\omega^{\alpha\mu\cdots\nu}) &=
\lambda\grad\alpha\Omega^{\alpha\mu'\cdots\nu'}\Pmm\cdots\Pnn
+
\lambda\inv\xi_{\alpha'}\Omega^{\alpha'\mu''\cdots\nu'}
\Pmm\cdots\Pnn\grad{\mu''}\xi^{\mu'}
\cr &\qquad\qquad
 + \cdots\ +
\lambda\inv\xi_{\alpha'}\Omega^{\alpha'\mu'\cdots\nu''}
\Pmm\cdots\Pnn\grad{\nu''}\xi^{\nu'}.
\cr}}
Since $\Omega$ is a form, one can insert the projection operators $\Pi$ to
replace the $\mu'\cdots\nu'$ contractions in the last terms (since there is
already one $\xi$ contracted with $\Omega$), and we get
\eqn\finlem{\eqalign{\gt_\alpha(\lambda\omega^{\alpha\mu\cdots\nu}) &=
\lambda\grad\alpha\Omega^{\alpha\mu'\cdots\nu'}\Pmm\cdots\Pnn \cr
&\qquad\qquad +
\lambda\inv\xi_{\alpha'}\Omega^{\alpha'\mu''\cdots\nu'}
\Pmm\cdots\Pnn\Pi^{\mu'''}_{\mu''}
\grad{\mu'''}\xi^{\mu'} + \cdots\cr}}
The second term and similar terms (represented by the dots) vanish by \xider\
 and we are left with
$${\gt_\alpha(\lambda\omega^{\alpha\mu\cdots\nu}) =
\lambda\Pmm\cdots\Pnn\grad{\alpha'}\Omega^{\alpha'\mu'\cdots\nu'}.}
\eqno(2.1)$$

\listrefs\bye